\documentclass[physics,article,accept,pdftex,moreauthors]{Definitions/mdpi} 
\firstpage{1} 
\pubvolume{1}
\issuenum{1}
\articlenumber{0}
\pubyear{2023}
\copyrightyear{2023}
\externaleditor{Academic Editor:}
\datereceived{ } 
\daterevised{ } 
\dateaccepted{ } 
\datepublished{ } 
\hreflink{https://doi.org/} 

\Title{Numerical Simulations of the Decaying Transverse Oscillations in the Cool Jet}

\TitleCitation{Numerical Simulations of the Decaying Transverse Oscillations in the Cool Jet}

\Author{Abhishek K. Srivastava * 
 and Balveer Singh}

\AuthorNames{Abhishek Kumar Srivastava and Balveer Singh}

\AuthorCitation{Srivastava, A.K.; Singh, B.}

\address[1]{
Department of Physics, Indian Institute of Technology (BHU), Varanasi 221005, India}

\corres{\hangafter=1 \hangindent=1.05em \hspace{-0.82em} Correspondence: asrivastava.app@iitbhu.ac.in 
}

\abstract{\textls[-5]{In the present paper, we describe a 2.5D 
 (two-and-a-half-dimensional) 
magnetohydrodynamic (MHD) simulation that provides a detailed picture of the evolution of cool jets triggered by initial vertical velocity 
perturbations in the solar chromosphere. We implement random multiple velocity, 
$V_{y}$, 
 pulses of amplitude 20--50 km s$^{-1}$ between 
1 Mm 
and 1.5 
Mm in the Sun's atmosphere below its transition region (TR). These pulses also consist of different switch-off periods between 50 
s 
and 300 s. The applied vertical velocity pulses create a series of magnetoacoustic shocks steepening above the TR. {These shocks interact} with 
each other in the inner corona, leading to complex localized velocity fields. The upward propagation of such perturbations creates low-pressure 
regions behind them, {which propel} a variety of cool jets and plasma flows {in the localized corona}.} The localized complex velocity fields generate 
transverse oscillations in some {of these} jets during their evolution. We study the transverse oscillations of a representative cool jet J$_{1}$, 
which moves up to the height of 6.2 Mm above the TR from its origin point. During its evolution, the plasma flows make the spine of jet J$_{1}$ {radially inhomogeneous}, 
which is visible 
in the density and Alfv\'en speed {smoothly} varying across the jet. The highly dense J$_{1}$, which {is} triggered along the 
significantly curved magnetic field lines, supports the propagating transverse wave of period 
 of approximately 
195 s with a phase speed of about \mbox{$
$125 km s $^{-1}$}. In the distance--time map of density, it is manifested as a transverse kink wave. However, the careful investigation of the distance--time maps of the x- and z-components of velocity reveals that these transverse waves are actually of mixed Alfv\'enic modes.
The transverse wave {shows evidence of damping} in the jet. We conclude that the cross-field structuring of the density and characteristic Alfv\'en speed within J$_{1}$ causes the onset of the resonant conversion and leakage of the wave energy outward to dissipate 
these transverse oscillations via resonant absorption. The wave energy flux is estimated as 
approximately of 
1.0
$\times$ 10$^{6}$ ergs cm$^{-2}$ s$^{-1}$. This energy, if it dissipates through the resonant absorption into the corona where the jet is propagated, is sufficient energy for the localized coronal heating.}


\begin{document}

\section{Introduction}

The Sun's chromosphere and its intricate magnetic structures possess numerous plasma ejecta, such as solar jets, spicules, mottles, surges, magnetic 
swirls, network jets, etc., at diverse spatio-temporal scales. These plasma ejecta may contribute to the transport of mass and substantial energy into the overlying layers of the Sun's atmosphere, i.e., the transition region (TR) and corona. The transport of mass and energy from the lower atmosphere to the outer corona may elucidate various physical mechanisms contributing to the localized heating, which is still an outstanding problem in solar physics (e.g., \cite{2000SoPh..196...79S, 2004Natur.430..536D, 2007Sci...318.1574D, 2010ApJ...722.1644S, 2011Natur.475..477M,2014Sci...346D.315D, 2015A&A...581A.131J, 2017NatSR...743147S, 2018NatAs...2..951S, 2018A&A...616A..99K, 2019AnGeo..37..891S, 2019NatCo..10.3504L, 2020ApJ...894..155S, 2020ApJ...897L...2P, 2021ApJ...913...59W, 2021ApJ...913...19M, 2021ApJ...918L..20H, 10.1093/mnras/stac252,2023NatAs.tmp..110Y} and references 
therein). 
 These chromospheric plasma ejecta are mostly collimated beam-like plasma structures. These jets may be associated with a variety of magnetohydrodynamic (MHD) waves, oscillations, and instabilities. Observations from space-borne and ground-based telescopes and their back-end instruments reveal that these jet-like structures are significant candidates accompanied by a range of physical processes that occur in the solar atmosphere (e.g., shocks, instabilities, heating, waves, flux emergence, and reconnection). Extensive studies reveal that such structures are ubiquitous and quasi-periodic on the limbs and disks of the solar atmosphere. These jet-like structures are also categorized based on their physical and morphological properties (e.g.,~\cite{2007Sci...318.1574D, 2014Sci...346D.315D, 2011Natur.475..477M, 2014Sci...346A.315T, 2015A&A...581A.131J, 2018NatAs...2..951S, 2019Sci...366..890S, 2019AnGeo..37..891S, 2021ApJ...913...19M, 10.1093/mnras/stac252} and 
references 
therein).

Recent high-resolution observations show that the chromospheric jet and jet-like structures (e.g., network jets, magnetic swirls, spicule-like rotating plasma structures, etc.) are triggered quasi-periodically, reaching various heights (2--12 Mm) in the solar atmosphere.~The typical lifetime of these jets is 3--12 min and their maximum velocity is found between 10 and 150 km s$^{-1}$ (e.g., \cite{2007Sci...318.1574D,2009SoPh..260...59P, 2014Sci...346D.315D} and 
 references 
therein).
 These jets are cool ($\simeq$10$^4$ K) and thin plasma structures ({width} 
300--1100 km), 
which disappear in the chromospheric lines 
before their downfall (e.g., \cite{2009SoPh..260...59P}).~This~effect may due to the fact that the jet's cool material is either rapidly expanding or some heating mechanism is at work within it \cite{10.1093/pasj/65.3.62}. A small-scale reconnection-generated chromospheric jet was also observed using He I 10830 \AA~and TiO 7057 \AA~high-resolution images as captured by the 1.6 m aperture Goode Solar Telescope at the Big Bear Solar Observatory, which revealed the thin dark threads inside the jet anchored in the quiet-Sun inter-granular lanes \citep{2021ApJ...913...59W}. The evolution of such cool jets and associated plasma processes provides a valuable insight into their triggering mechanisms in the Sun's atmosphere. 
Various single- and two-fluid MHD models provide a detailed picture of the evolution and triggering mechanisms of these jets, as well as the physics of their various drivers in the lower solar atmosphere. These models are primarily based on the evolution of the perturbation in the form of gas pressure and/or velocity pulses that interact with the granular motion. Such interactions and the complex structuring of the magnetic fields may also generate the magnetic reconnection at small spatial scales, which may cause localized heating and the formation of jet-like plasma ejecta in the lower solar atmosphere (e.g., \cite{1992PASJ...44L.173S,1995Natur.375...42Y, 2004Natur.430..536D, 2007ApJ...666.1277H,2008ApJ...683L..83N,2010A&A...519A...8M,2011A&A...535A..58M, 2013ApJ...770L...3K,2018A&A...616A..99K, 2015A&A...581A.131J,2017ApJ...848...38I,2017ApJ...849...78K, 2019AnGeo..37..891S, 2021ApJ...918L..20H,2021MNRAS.505...50G} and 
 references 
therein). 
{In conclusion, there is a non-exhaustive list of studies where the evolution, triggering, kinematics, and energetics of the various solar jets are mentioned, highlighting the significance of these studies (e.g.,~\cite{2012ApJ...750...51K,2014A&A...566A..90M,2014Ap&SS.352....7T,2015A&A...573A.130P,2018ApJ...860..116M} and 
 references 
therein).}\enlargethispage{0.5cm}
  
Several extensive observations reveal that the inhomogeneous and complex motion of the plasma produces various wave modes and instabilities in the jets \citep{2021JGRA..12629097S}. These modes of oscillation and MHD instabilities are generated in situ either by 
 quite 
complex plasma motions near the regions of concentration of the field lines  or by the shearing motion in the ambient plasma of the jet’s spire. The jet-like structures may oscillate along the field lines with periods of several minutes (e.g., \cite{1983SoPh...88...35K,2009SSRv..149..355Z, 2003ApJ...595L..63D, 2005A&A...438.1115X, 2016A&A...585L...6P}). The oscillations across the axis of the jet have been reported 
 in Refs.
\cite{article,Jess_2011,2016A&A...585L...6P}. These transversal oscillations show periodic behavior due to the generation of the kink waves. The propagation of the kink waves in the jet-like structures can be observed in the {$H\alpha$} spectral line (e.g., \cite{Jess_2011}). 
  Apart from jet-like structures, transversal kink oscillations are also highly ubiquitous in solar coronal loops, and 
 these kink motions 
can also be triggered by the motion of the jet and flaring disturbances 
\cite{2016SoPh..291.3269S,2016A&A...585L...6P,2020A&A...638A..32Z,2021A&A...646A..12D,2021SSRv..217...73N,2022ApJ...937L..21Z}. However, there are several controversies in the interpretation of transverse oscillations in the jet-like structures.~They~may be pure Alfv\'en waves, kink waves or Alfv\'enic waves, appearing as a transverse motions in sufficiently resolved observational imaging or velocity data.~The MHD waves, which have a wave mode corresponding to $m=1$, are called kink oscillations, while Alfv\'nic waves are described as a result of predominantly a mixed mode in non-homogeneous waveguides \cite{2021JGRA..12629097S}. The linear kink waves are incompressible in nature, and they exhibit periodic displacement resulting from magnetic tension forces (e.g., \cite{2009A&A...503..213G}). These transverse kink oscillations are ubiquitous in the jet-like structures, and they have been found to be mainly due to the overshooting of the complex and shearing motion of the plasma as rebound shocks and the convective motion of the photospheric plasma (e.g., \cite{1988ApJ...327..950S,2008POBeo..84..507V}). It is well known that lower solar atmospheric oscillations are susceptible to features of mode coupling, in which one magnetoacoustic wave mode is coupled to another (slow or fast magnetoacoustic waves) and the wave energy can be transferred to another mode in the particular atmospheric condition (e.g., \cite{2005LRSP....2....3N}). Apart from this, kink oscillations are ubiquitously detected in a variety of coronal and chromospheric structures, and their {damping} is mostly governed by the processes (e.g., resonant absorption) arising due to the transverse inhomogeneities  of the plasma (e.g., density),  the magnetic field, and the characteristic velocities across the flux tube \cite{2002A&A...394L..39G,2006RSPTA.364..433G,2021JGRA..12629097S,2021SSRv..217...73N}.

In this paper, we provide a physical scenario of the evolution of a variety of cool jets and flows under the influence of random vertical velocity perturbations in 
the chromosphere, creating a complex velocity field. A special emphasis is further placed on the study of a cool jet possessing  decaying transverse 
oscillations. We perform a 2.5D 
(two-and-a-half-dimensional)  
numerical simulation of non-linear MHD equations for a single-fluid plasma with the implementation of vertical velocity pulses around the null point configuration of the magnetic field. The evolution of the transversely oscillating jet in our  model matches the peculiar observational signature of the kink oscillations in the cool jet-like structure reported in previous literature (e.g.,~\cite{Jess_2011}). We analyze the triggering mechanism of these transverse oscillations, their properties, and the most likely cause of their {damping} in the model jet.~{The possibility of launching multiple jets due to random velocity fluctuations and the subsequent evolution of kink motion and its damping through resonant absorption is an unique concept described in the present work.} The paper is structured as follows. Section \ref{sec:2} provides the description about fundamental MHD model of the solar atmosphere, as well as its initial equilibrium configuration and perturbations. Section \ref{sec:3} presents the detailed scientific results associated with numerical simulations. Section \ref{sec:4} presents the discussion and conclusions. 

\section{MHD Model of Cool Jets and Numerical Methods}\label{sec:2}
\subsection{The Ideal MHD System}

We consider an ideal MHD system to model the onset of various cool plasma ejecta and waves in the magnetized and gravitationally stratified Sun's 
atmosphere.~In~this model, we do not consider non-ideal effects such as dissipative effects, resistivity, magnetic diffusivity, or heating and/or 
cooling effects of the plasma, and also avoid the initial velocity of the plasma particles. In the present paper, we utilize MHD approximation to 
understand the evolution of various cool jets on both sides of a null point geometry as modelled in the solar atmosphere. 
The governing equations of the model are 
(e.g., \cite{2007ApJS..170..228M,2020ApJ...894..155S, 10.1093/mnras/stac252}): 
\begin{equation}
    \frac{\partial\varrho}{\partial t} + \nabla\cdot(\varrho\textbf{V}) = 0, 
\end{equation}
\begin{equation}
     \varrho\frac{\partial\textbf{V}}{\partial t} +\varrho(\textbf{V}\cdot\nabla)\textbf{V} = -\nabla p + \frac{1}{\mu}(\nabla\times\textbf{B})\times\textbf{B} +\varrho\textbf{g},
\label{eq2}
\end{equation}
\begin{equation}
  \frac{\partial\textbf{B}}{\partial t} = \nabla \times (\textbf{V} \times\textbf{B}),\hspace{1cm}    \nabla\cdot\textbf{B} = 0,  
\end{equation}
\begin{equation}
  \frac{\partial p}{\partial t} + \nabla \cdot (p\textbf{V}) = (1- \gamma) p \nabla\cdot\textbf{V},  
\end{equation}
\begin{equation}
    p= 
 k_{B} \frac{\varrho T}{m}.
\end{equation}

The variable $\varrho$ 
represents the mass density, $\textbf{V}
$ 
represents the plasma velocity, $t$ denotes the time, 
$T$ represents the temperature, 
and variable $p$ represents the thermal pressure. $ \mu$ represents the magnetic permeability, $\textbf{B}$ represents the magnetic field, and $\textbf{g}$ represents the gravitational acceleration. We take the value of $\textbf{g}$ equal to 
 27,400 
cm s$^{-2}$, which is an average value at the Sun's surface (i.e., at the photosphere). 
$k_{B}$ 
represents the Boltzmann constant and $ m $ denotes the mean mass of the plasma particles in the simulation domain.
 We~take the value of $ m $ equal to 1.24 $m_{H}$, where $m_{H}$ is the hydrogen mass.~$\gamma$ represents the specific heat ratio, which is equal to 
5/3 for the three degrees of~freedom.\enlargethispage{0.5cm}

We implement the realistic temperature model, which was derived from {{Avrett} and \linebreak {Loeser}~\cite{2008ApJS..175..229A}} \textls[-5]{by considering the line profiles {as observed in} the Sun’s atmosphere.~The background physical parameters, such as density, pressure, etc., are estimated {taking into account} this realistic temperature profile and the initial hydrostatic equilibrium {condition}. Therefore, the stratified solar atmosphere is described here as a realistic  model atmosphere (\mbox{
see 
Section \ref{S-Equli}}). We study {only} the formation mechanism of the jets and evolved oscillations present in the longer cool jets in the ideal MHD regime.  Further, using 
2.5D 
simulation, we show that as the evolved transverse oscillations {dampen} on the body of the jet, simultaneously, the 
$V_z$ 
velocity signals remain and are enhanced, which is the signature of the excitation of Alfv\'enic perturbations 
 (see 
Section \ref{sec:3}). This 
 well 
 demonstrates the efficient resonant absorption mechanism that leads to the quick {damping} of transverse kink oscillations. We do not aim to study the {morphological and} kinematical properties (e.g., length, width, lifetime, and other features) of these jets. We also neglect the non-ideal classical dissipative terms, such as  resistivity, magnetic diffusivity, and heating or cooling, which may not significantly affect the evolution and {damping} of the oscillations present in the jets as efficiently as resonant absorption does in the present simulation. The present simulations aim to demonstrate that resonant absorption is set after some time (300--400 s) of the evolution of complex transverse motions in the jet, which further {dampen} the transverse waves inside the jet by transferring energy to the boundary of the jet in the form of surface Alfv\'enic perturbations. We present these results in Section \ref{sec:3} in a detailed manner.}

\subsection{Equilibrium Condition of the Model Solar Atmosphere} 
      \label{S-Equli}    
      
Initially, there is no plasma movement in the considered model of the solar atmosphere (i.e., the speed of the background plasma flow, 
$V_{e} = 0)$. This variable represents the plasma flow in external background atmosphere. The 'e' subscript refers to the external medium. 
Therefore, it is implemented from the photosphere to the inner corona (i.e.,~40~Mm) in its static equilibrium. This model 
atmosphere 
is supported by a force- and current-free medium. Therefore, the configuration of the magnetic field is initially potential and can be expressed as follows (e.g., \citep{2019AnGeo..37..891S, 2020ApJ...894..155S, 10.1093/mnras/stac252}):
\begin{equation}
\nabla \times \textbf{B} = \textbf{0} , \hspace{2cm} (\nabla \times \textbf{B}) \times \textbf{B} = \textbf{0}.
\end{equation} 

Our model solar atmosphere is made as current-free and force-free initially. Here, 
gradient \textbf{B} 
is not equal to zero; however, it has negligible values. Initially, we consider all the background velocities as zero, and the force due to pressure 
gradient is balanced by {the} gravity force.  Therefore, 
Equation 
 (\ref{eq2}) 
(momentum equation) 
 indeed  
shows that if these conditions are applied, 
i.e., 
 $\textbf{V} = 0,\; \varrho \textbf{g} = -\nabla p$,  
the Lorentz  force is nearly equal to zero. This approximation is referred to as the current-free condition that holds 
the model solar atmosphere {in the initial equilibrium}. {Therefore, this approximation 
 shows 
that all the forces are in equilibrium. This condition is referred to as the force-free condition of the model solar atmosphere.

 The equilibrium configuration of the magnetic field, \textbf{B}, 
with the null point geometry is estimated as follows 
(e.g., 
\cite{1985ApJ...293...31L}):
\begin{equation}
\begin{split}
B_{x} = \frac{-2S_{1}(x-a_{1})(y-a_{2})}{[(x- a_{1})^2 + (y - a_{2})^2]^2} \\ +\, 
\frac{-2S_{2}(x-b_{1})(y-b_{2})}{[(x- b_{1})^2 + (y - b_{2})^2]^2} \\ +\, 
   \frac{-2S_{3}(x-c_{1})(y-c_{2})}{[(x- c_{1})^2 + (y - c_{2})^2]^2} \\ +\, 
   \frac{-2S_{4}(x-d_{1})(y-d_{2})}{[(x- d_{1})^2 + (y - d_{2})^2]^2},
   \end{split}
\end{equation} 
\begin{equation} 
\begin{split}
B_{y} = \frac{2S_{1}(x-a_{1})^2 - S_{1}((x-a_{1})^2 + (y-a_{2})^2)}{[(x- a_{1})^2 + (y - a_{2}]^2}\\ +\, 
\frac{2S_{2}(x-b_{1})^2 - S_{2}((x-b_{1})^2 + (y-b_{2})^2)}{[(x- b_{1})^2 + (y - b_{2})^2]^2}
 \\ +\,  
 \frac{2S_{3}(x-c_{1})^2 - S_{3}((x-c_{1})^2 + (y-c_{2})^2)}{[(x- c_{1})^2 + (y - c_{2})^2]^2} \\ +\, 
 \frac{2S_{4}(x-d_{1})^2 - S_{4}((x-d_{1})^2 + (y-d_{2})^2)}{[(x- d_{1})^2 + (y - d_{2})^2]^2},
\end{split}
\end{equation} 
\begin{equation}
B_{z} \ne 0.
\end{equation} 

Here, $B_{x}$ and $B_{y}$ represent the magnetic field components respectively in the horizontal (i.e., X-axis) and vertical (i.e., Y-axis) directions. There is an implementation of the constant magnetic field in the line of sight direction, i.e., $B_{z}$ = 5 G. There are charges $S_{1}$, $S_{2}$, $S_{3}$, and $S_{4}$. $S_{1}$ has coordinates ($a_{1}$, $a_{2}$), $S_{2}$ has coordinates ($b_{1}$, $b_{2}$), $S_{3}$ has coordinates ($c_{1}$, $c_{2}$), and $S_{4}$ has coordinates ($d_{1}$, $d_{2}$). 
We take them as $ S_{1}=S_{3}$ = $-$56 G and $S_{2}=S_{4}$ = 1120 G. 
We fix the horizontal positions of the poles at $a_{1}$ = 
$-$10 
 Mm, 
$b_{1}$ = $-$5 
 Mm, $c_{1}$ = 5 
 Mm, 
and $d_{1}$ = 10 
 Mm on the X-axis. We also fix the vertical positions of the poles at $a_{2}$ = $c_{2}$ = $-$10 
 Mm, $b_{2}$ 
=  $d_{2}$ = $-$5 
 Mm below the bottom boundary of the solar photosphere (i.e., 0 
 Mm). 
Therefore, the magnetic field lines originate from the convection zone. The magnetic configuration displays the null point geometry of the magnetic field lines (see Figure~\ref{Figure:1}). The field lines on both of its sides are curved fields smoothly extending up to 40 Mm in the vertical direction, as visualized in the model solar atmosphere (see Figure~\ref{Figure:1}, 
left). 
Some of the low-lying loop structures are also seen, 
filling the lower corona and chromosphere. 
 0 
Mm, which represents the bottom of the solar photosphere. We launch the pulses above the solar photosphere, so there are no dynamics occurring below the photosphere boundary. We took the position and strength of the magnetic fields of the poles such that we can fix the location of the null point in the model solar atmosphere.   
\begin{figure}[H]   
\centerline{\includegraphics[width=0.5\linewidth]{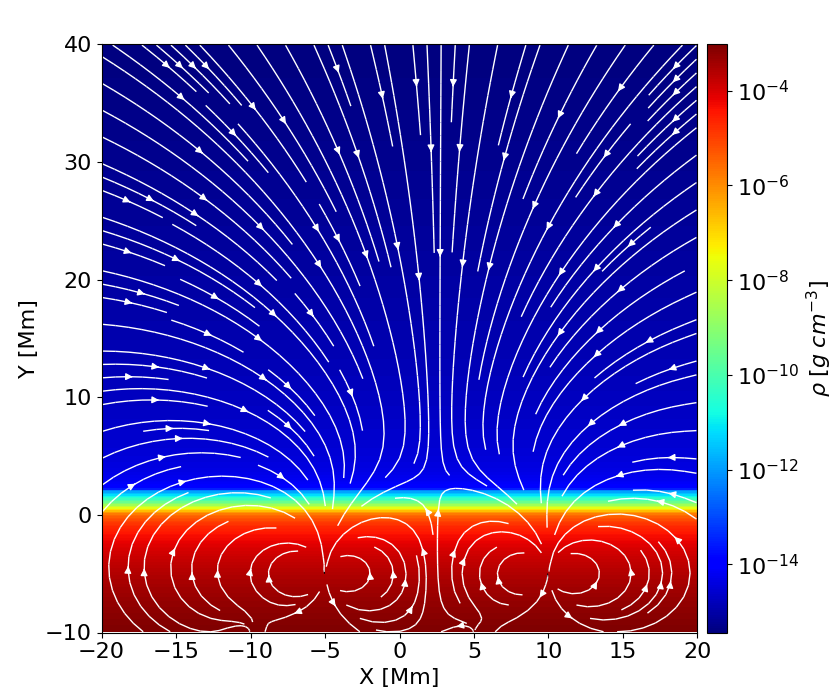}
  \includegraphics[width=0.5\linewidth]{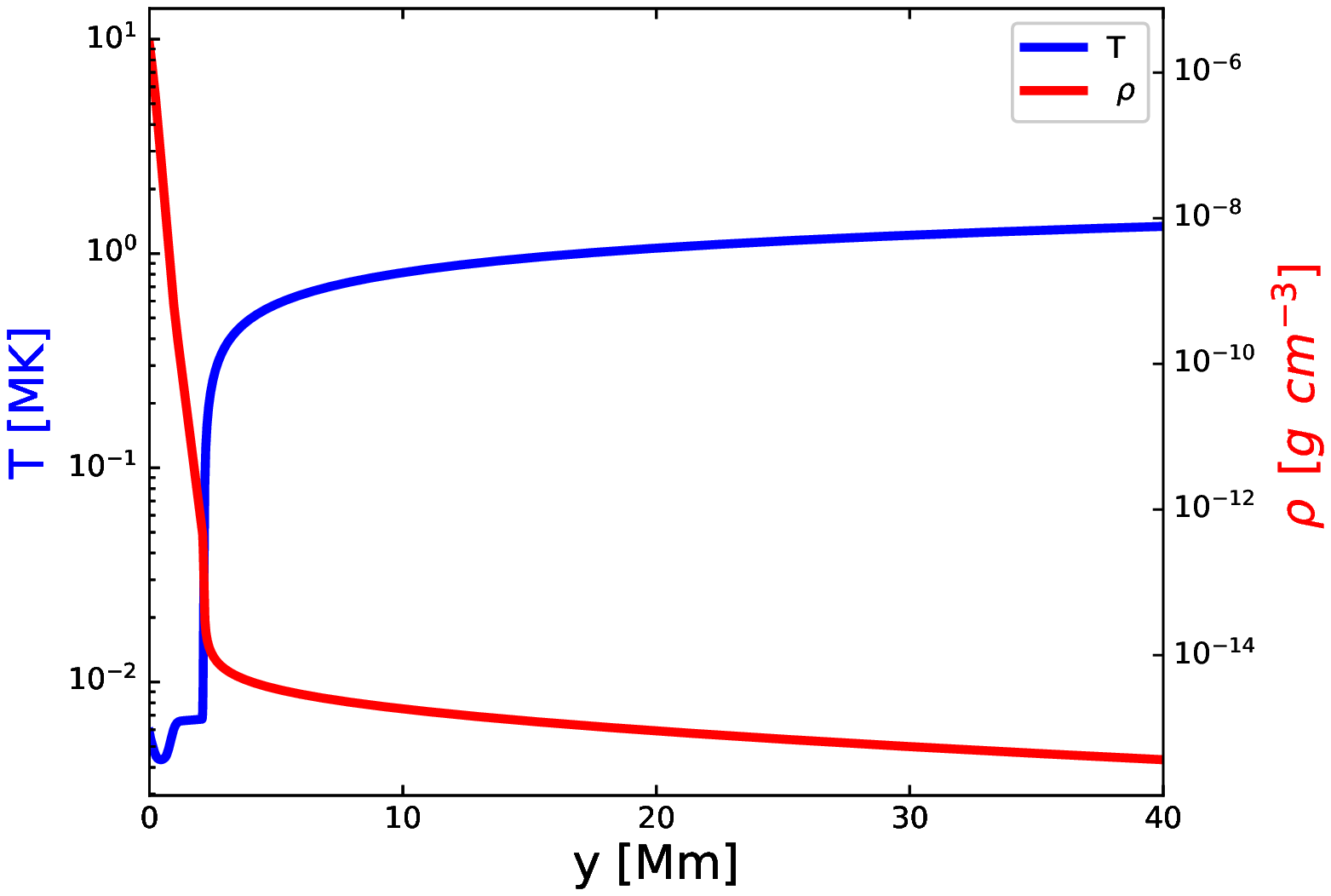}}
\caption{\textbf{Left:} 
the equilibrium magnetic field lines in the stratified model solar atmosphere. The null point is formed above the transition region (TR) 
 in the lower corona. 
\textbf{Right:} 
the variations in temperature, $T$ 
(blue curve), and mass density, 
$\varrho$ 
 (red curve),  with respect to the vertical direction, 
$y$, 
 in the model solar 
atmosphere.}\label{Figure:1}
\end{figure}


We consider a realistic and gravitationally stratified model solar atmosphere that initially follows the physical scenario of hydrostatic equilibrium. In the model solar atmosphere, the pressure gradient and gravity force are balanced by each other, resulting in the medium being approximately force-free initially. Therefore, this is expressed as follows (e.g., \cite{2019AnGeo..37..891S, 2020ApJ...894..155S, 10.1093/mnras/stac252}): 
\begin{equation}\\ \\
- \bigtriangledown p + \varrho \textbf{g} = 0.
\end{equation}

We express the equilibrium configuration of mass density and thermal pressure in the longitudinally structured solar atmosphere, which maintains hydrostatic equilibrium initially. The thermal pressure ($p$) and mass density ($\varrho$) with respect to the vertical 
direction, 
$y$, 
are determined as follows 
(e.g.,~\cite{2019AnGeo..37..891S, 2020ApJ...894..155S, 10.1093/mnras/stac252}):
\begin{equation}\\ \\
p_{e}(y) = p_{0}\,\exp\left(- \int_{y_{0}}^{y}\frac{dy'}{\Lambda(y')}\right),
\label{eq11}
\end{equation}
and 
\begin{equation}
\varrho_{e}(y) 
= \frac{p_{e}(y)}{g\Lambda(y)},\;\; \\ \\
{\rm where}\;\;  
\Lambda(y) =\frac{k_{B} T_{e}(y)}{mg}.
\label{eq12}
\end{equation}

Here, $p_{0}$ denotes the thermal pressure at the reference level $y_{0}$ = 10 Mm. The 
subscript {\it 'e'} 
 indicates that the associated variables are in equilibrium. 

\textls[-5]{We consider the magnetized and gravitationally stratified solar atmosphere, in which all background physical parameters are structured 
vertically. The temperature 
$T_e{(y)}$} 
variation with respect to the vertical direction ($y$) 
in the realistic model solar atmosphere is illustrated in Figure~\ref{Figure:1}, right
(blue curve).} This temperature profile was calculated 
by analyzing the observed line profiles in the solar atmosphere in \citet{2008ApJS..175..229A}.~The~typical value of the temperature is $\simeq$ 5807 
K at height $y = 0$ 
Mm, which corresponds to the temperature of the solar surface. On moving toward a higher altitude, it gradually decreases to $\simeq$4350~K at $y = 0.5$~Mm, which indicates the temperature minimum region at the top of the solar photosphere as well as bottom of the chromosphere. In the present temperature profile, the solar transition region is located at altitude $y$ $\simeq$ 2.1 Mm, where the temperature rises sharply and reaches $\simeq$100,931 K. The equilibrium profile of mass density with respect to 
height, 
$y$, 
is shown in 
Figure~\ref{Figure:1}, right (red curve).
 This profile is expressed by 
Equations~(\ref{eq11}) and (\ref{eq12}). 
The value of mass density at $y$ = 0 
Mm (i.e., the solar photosphere) is $\simeq$$10^{-6}$ g cm$^{-3}$, while, at $y$ = 0.5 Mm (the solar chromosphere), it attains a value of \mbox{$10^{-8}$ g cm$^{-3}$}. When we move toward higher altitudes, the value of the mass density smoothly falls and reaches $\simeq$ $10^{-14}$ g cm$^{-3}$ at the TR, i.e., $y$ = 2.1 Mm in our model~atmosphere. 

\subsection{Numerical Methods}

We use the PLUTO code to solve the set of single-fluid MHD equations in the ideal regime of the plasma numerically (e.g., \cite{2007ApJS..170..228M}). It solves the set of non-linear MHD equations by implementing the initial as well as boundary conditions. The initial setup for the model solar atmosphere is chosen to be responsible for the generation of cool jets moving from the chromosphere upward along the field lines. 
In this model, the time integration is estimated using the third-order Runge--Kutta method. We apply the Roe solver to solve the flux calculation using decomposition of the Roe matrix of the characteristics 
 (see 
\cite{2007ApJS..170..228M}). We fix the Courant--Friedrichs--Lewy number at 0.25. The details of the numerical setup have been described previously (e.g., \cite{2019AnGeo..37..891S, 2020ApJ...894..155S, 10.1093/mnras/stac252}). 

We fix the model solar atmosphere with a horizontal span, 
$x$, 
of 40 Mm ($-$20, 20) and a vertical span,  
$y$, 
of 40 Mm (0, 40).~
In our model, the horizontal span is equally divided into 2560 cells as a uniform grid.~We cover a vertical span via uniform and stretched grids, 
equally divided into 320 cells from 0 Mm 
to 5 
 Mm 
and 1120 cells from 5 Mm 
to 40~Mm. 
Therefore, the spatial resolution is 15.625 km per numerical cell from $y=0$ 
Mm to $y=5$ 
Mm, and 31.25 Km per numerical cell from $y=5$ 
Mm and above in the model solar atmosphere. 
The vertical velocity signals generated by the velocity perturbations are 
 significantly  
weak at the top zone in the simulation box. Therefore, the incoming signal is diffused near the top zone boundary due to  the stretched grids, which have been already set in the overlying region of the simulation domain. The incoming signals have no discernible effect on jet triggering during the transition from uniform grids to stretched grids. 
In our model, we apply the min-mod limiter, which 
 turns to be successful 
from a diffusion perspective by reducing/restricting the reflection of the dynamics from the top zone in the model solar atmosphere (e.g., \cite{10.1093/mnras/stac252}). However, there are no substantial dynamics occurring in the top zone of the simulation box that would have an effect on the region of interest. Our region of interest is located around the transition region in our model solar atmosphere. 
We run the simulation with eight processors using a multiple passage interface (MPI) up to 1800 s and save the output data file at every 5 s. 

\textls[-5]{{We apply all four boundaries of the simulation box in such a way that the boundary maintains the equilibrium values for all the plasma quantities there.} We establish user-defined boundaries, which is the simplest approach to put forward implementing
appropriate 
boundaries in the presence of gravity in our gravitationally stratified  model solar atmosphere. The details of the imposed boundary condition have been described in various works (e.g., \cite{2019AnGeo..37..891S, 2020ApJ...894..155S, 10.1093/mnras/stac252}).
}
\subsection{Perturbation}\label{section:2.4} 

Initially, we apply 30 randomly generated vertical velocity pulses in the chromosphere between 1 
and 1.5 Mm height, which is accountable for the triggering of cool jets moving up in the model solar atmosphere along the magnetic field lines. These vertical velocity perturbations are applied for a fixed time ranging from 50 to 300 s (e.g., \cite{2021ApJ...913...19M}). The applied perturbation is represented by the following equation (e.g., \cite{2021ApJ...913...19M}): 
 \vspace{12pt}
\begin{equation}\\ 
\begin{split}
V_{y} = A_{v} \times \exp \left(-\frac{(x - x_{0})^2 + (y - y_{0})^2}{w^2}\right) \\ 
\times 
\left[\tanh \left (\frac{\pi (t-P)}{P} + \pi \right) + 1
\right]. 
\end{split}
\label{eq13}
\end{equation} 

Here, $A_{v}$ represents the amplitude of the velocity pulse, while the pair 
 $ x_{0}$ 
 and   
 $y_{0} $ 
represent the horizontal ($x$) and vertical ($y$) positions of the pulses. 
 $w$ 
denotes the full width at 
half 
maximum (FWHM) of the applied Gaussian pulse and  
  $P$ 
represents the period of the pulses. In the present model, we apply 30 random vertical velocity pulses of 
amplitude ranging between 20 and 50 km s$^{-1}$ (e.g., \cite{2021ApJ...913...19M}). We also choose the value of $w$ to be between 50 and 100 km. These 
pulses are applied  between $-$10 Mm 
and 10 Mm on the horizontal span and between 1 Mm 
and 1.5 Mm on the vertical span of the model solar atmosphere. The above expression of the Gaussian pulse is with a hyperbolic tangent, which is used 
to switch off the phase of the perturbation for a defined switch-off period. We select the switch-off period (P) of the pulses between 50 s  
and 300 s (e.g., \cite{2021ApJ...913...19M}). These strong vertical velocity pulses generate a variety of cool jets, in which some of them exhibit 
transverse oscillations.

\textls[-10]{In our model solar atmosphere, the medium is approximated as force-free and current-free. The solar photosphere is dominated by plasma pressure over the magnetic pressure. Therefore, the solar photosphere is not fully force-free. In this condition, if we implement the velocity field at the solar photosphere, the plasma dominates over the magnetic and photospheric plasma and does not obtain enough energy to propagate as a jet/spicule-like  structure along with the open magnetic field lines of the solar atmosphere. We implement the velocity field at the chromospheric height, at which the magnetic pressure is nearly equal to the plasma pressure, i.e., plasma beta is close to one.  In this condition, jet/spicule-like plasma motions occur close to the observations of these chromospheric structures. Therefore, photospheric velocity fields are not more appropriate, such as chromospheric velocity fields. Moreover, we seek to mimic the family of cool jets (not specifically the typical spicules only), and we describe them as cool spicule-like jets (e.g., network jets, macrospicules, etc.) that originate in the solar chromosphere in the form of chromospheric plasma ejecta. {The chromosphere is subject to impulsive energy releases at small spatial scales, and even the photospheric tiny magnetic flux cancellations may lead to chromospheric reconnections; eventually, the plasma outflows in the form of cool jets (e.g.,~\cite{2000A&A...360..351W,2007ASPC..369..243C,2019Sci...366..890S,2019ApJ...873...79C,2023EPJP..138..209S} and 
 references 
therein).}}

\section{Results}\label{sec:3} 
 
  The random vertical velocity pulses launched in the solar chromosphere (as per Equation (\ref{eq13});  
 see 
Section \ref{section:2.4}) create a series of field-aligned magnetoacoustic  perturbations in the solar atmosphere that possess a typical magnetic structuring 
(see 
Section \ref{S-Equli}). Essentially, even if there is one such pulse that evolves in the solar chromosphere, it will generate a slow magnetoacoustic shock while moving up into the stratified atmosphere. This will  create a low-pressure region, which further causes the movement of cool plasma in the upward direction, forming a thin, cool spicule-like jet \cite{2010A&A...519A...8M,2011A&A...535A..58M}.
  
\begin{figure}[H]   
\includegraphics[width=12cm,height=8cm]{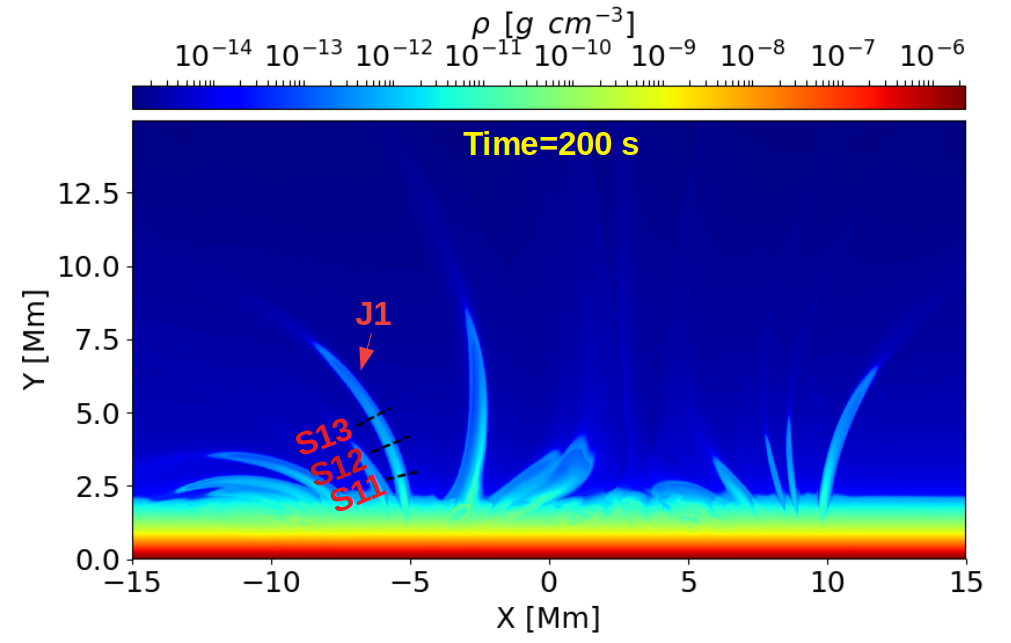}
\caption{The mass density, $\varrho$, 
map 
shown with three dotted slits drawn on the cool jet ` J$_{1}$'. 
 One sees 
the distance--time diagrams over these chosen slits. The 
 video \cite{video}  
 associated with the shown 
field-of-view 
(FOV) 
 demonstrates that multiple velocity perturbations in the chromosphere, once steepened in the inner corona, interact 
with each other and generate episodic complex velocity fields locally 
(see 
the dynamics/motion of the overlaid velocity vector arrows in the video). Some of the evolving magneto-plasma structures (here, spicule-like jets) exhibit transversal oscillations under the influence of the evolved complex velocity fields.}
\label{fig}
\end{figure}

  In a similar manner, when multiple such pulses initially propagate from the solar chromosphere to the TR and thereafter in the inner corona, they steepen, forming multiple field-aligned shocks, and cause the evolution of a variety of cool jets, as shown in Figure~\ref{fig}. 
The video \cite{video} 
 associated with the shown 
 field-of-view 
(FOV) 
(see 
Figure~\ref{fig}) demonstrates that once multiple velocity perturbations in the chromosphere are steepened in the inner corona, they interact with each other and generate episodic complex velocity fields locally. The dynamics of the plasma and overlaid velocity vectors in the 
video 
 well 
demonstrate the evolution of this complex physical scenario in the localized corona. The evolving magneto-plasma structures (here, spicule-like jets) exhibit  transversal oscillations under the influence of this. These spicule-like cool jets rise and fall quasi-periodically, as recently reported by \citet{10.1093/mnras/stac252}. In this paper, we focus on the dynamics of a representative spicule-like cool 
jet, 'J$_{1}$', 
as displayed in Figure~\ref{fig}. The details of the plasma dynamics at the null point are beyond the scope of this paper and will be discussed separately in a 
forthcoming studies. 
  
\begin{figure}[H]   
\mbox{
\includegraphics[width=4cm,height=6cm]{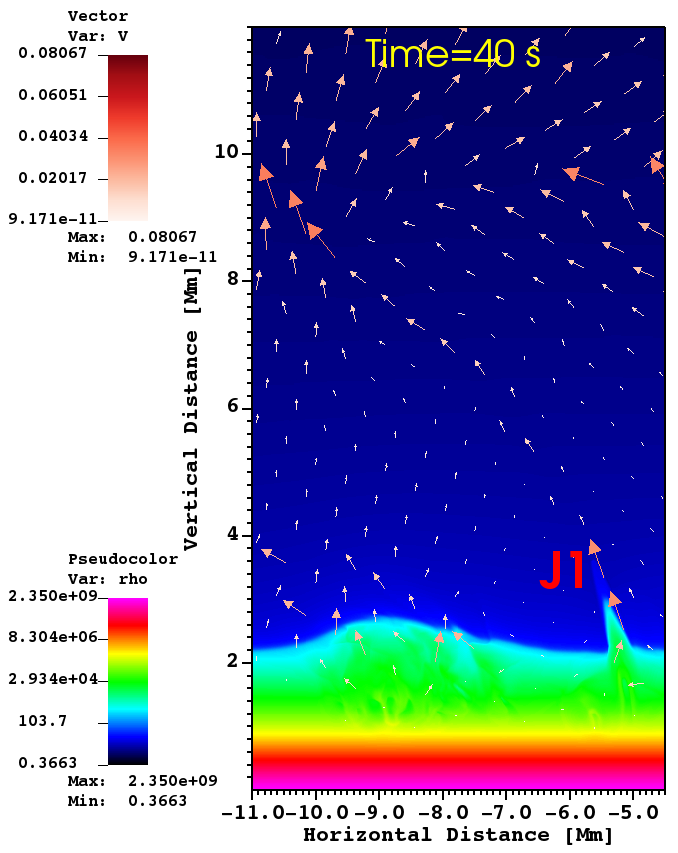}
\includegraphics[width=4cm,height=6cm]{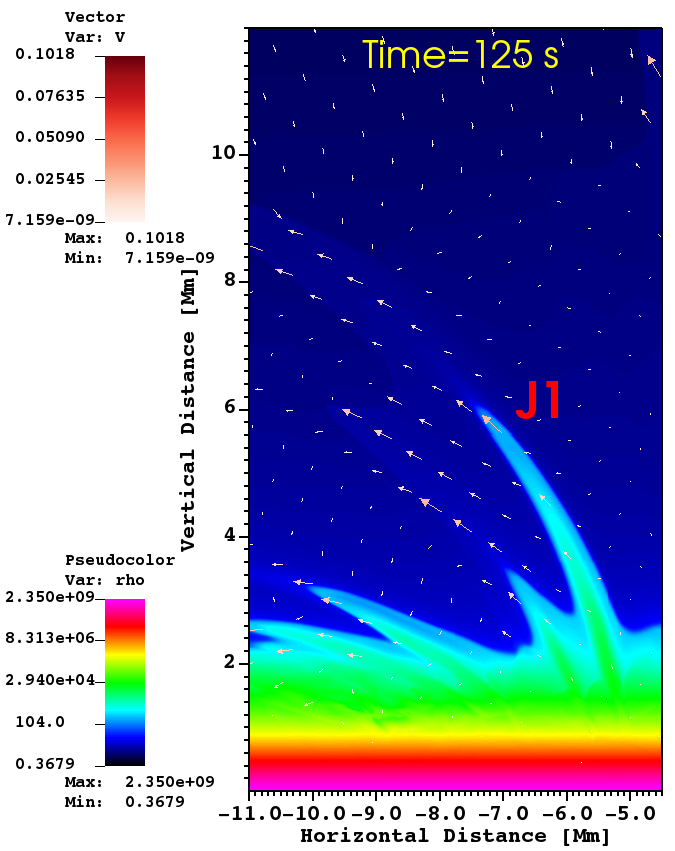}
\includegraphics[width=4cm,height=6cm]{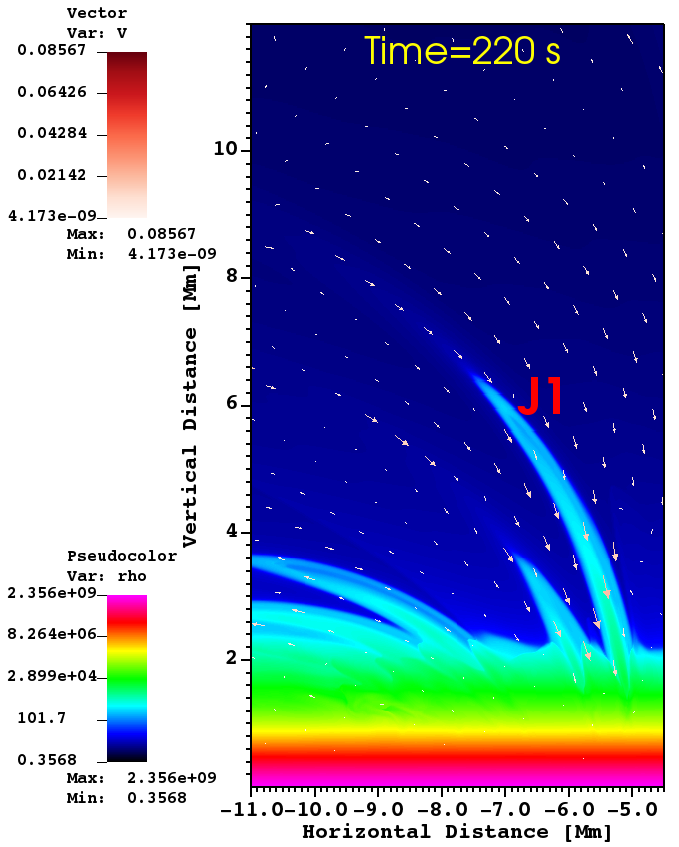}
}
\mbox{
\includegraphics[width=4cm,height=6cm]{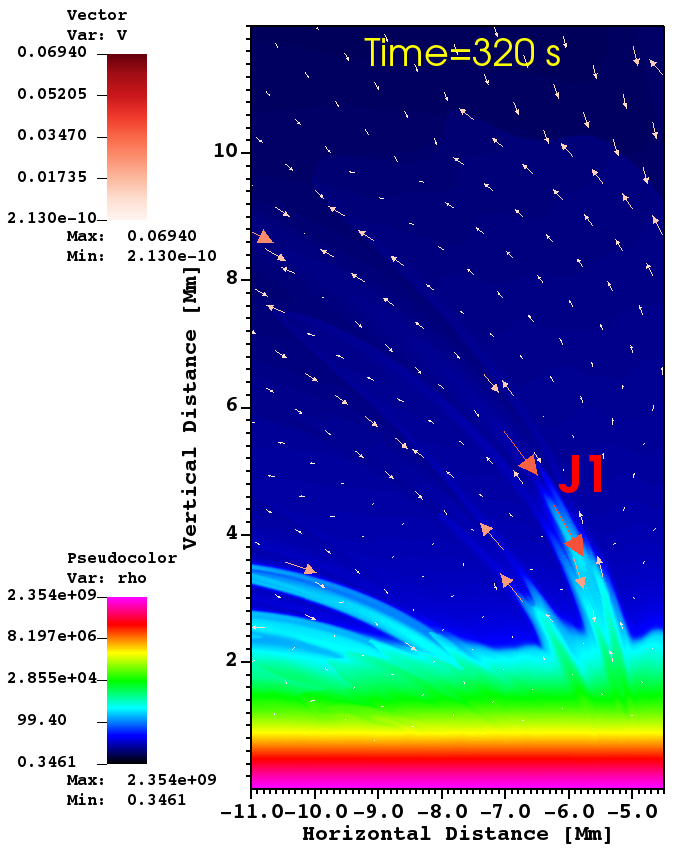}
\includegraphics[width=4cm,height=6cm]{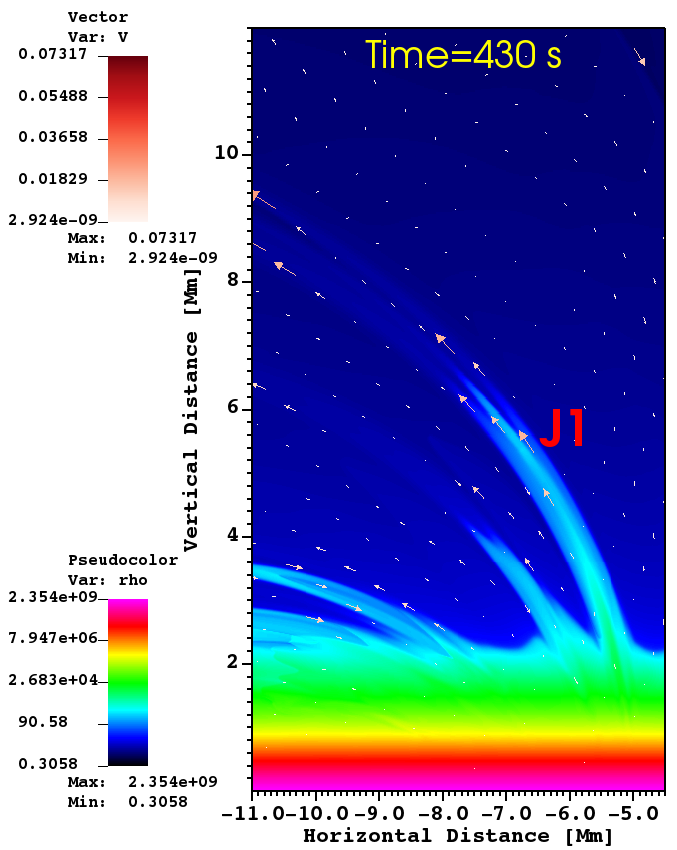}
\includegraphics[width=4cm,height=6cm]{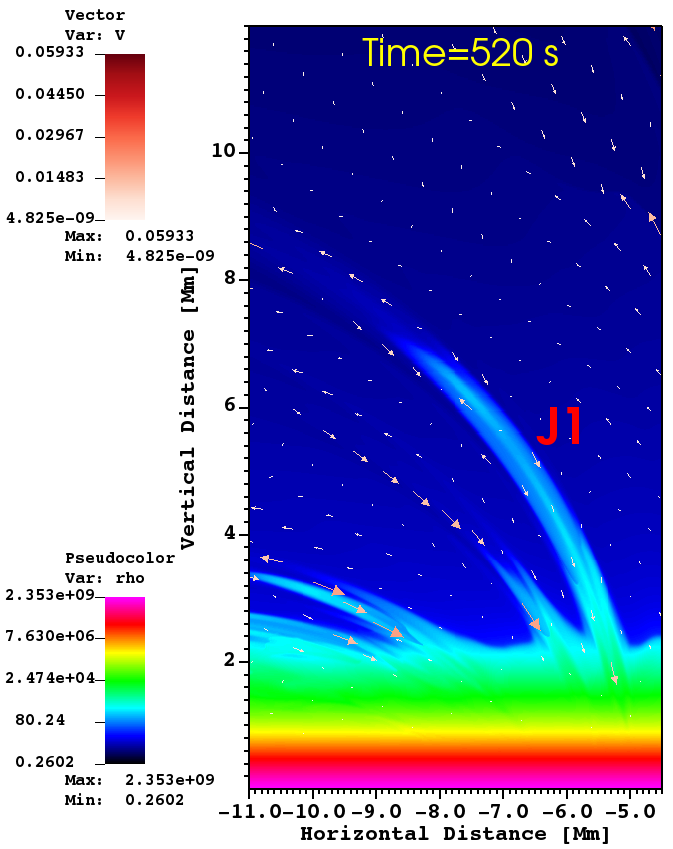}
}
\mbox{
\includegraphics[width=4cm,height=6cm]{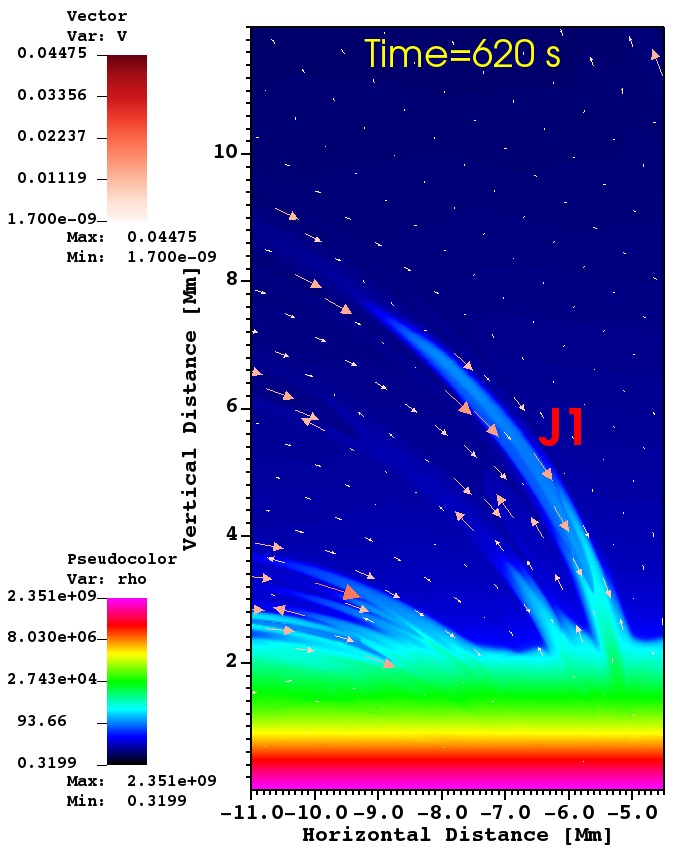}
\includegraphics[width=4cm,height=6cm]{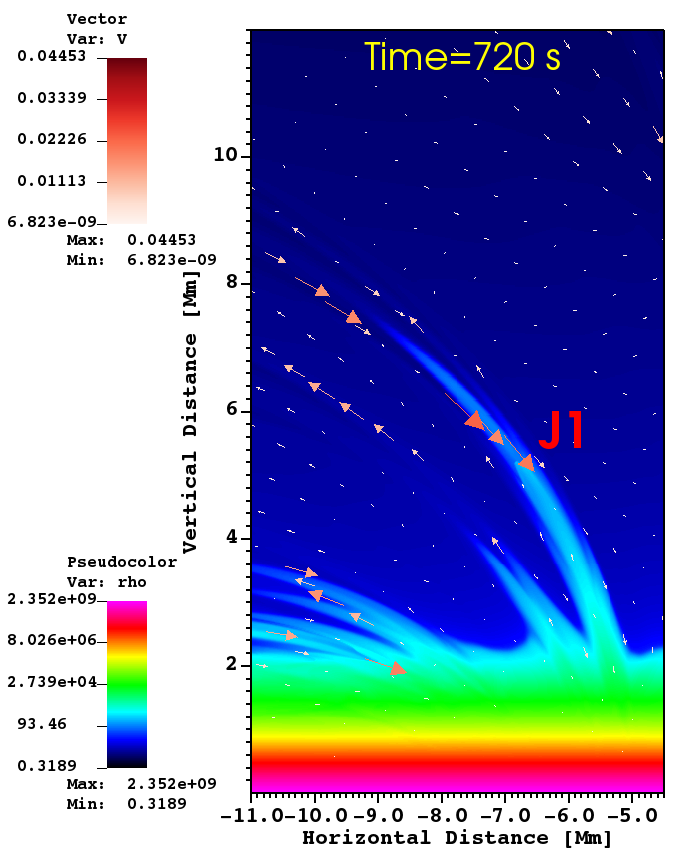}
\includegraphics[width=4cm,height=6cm]{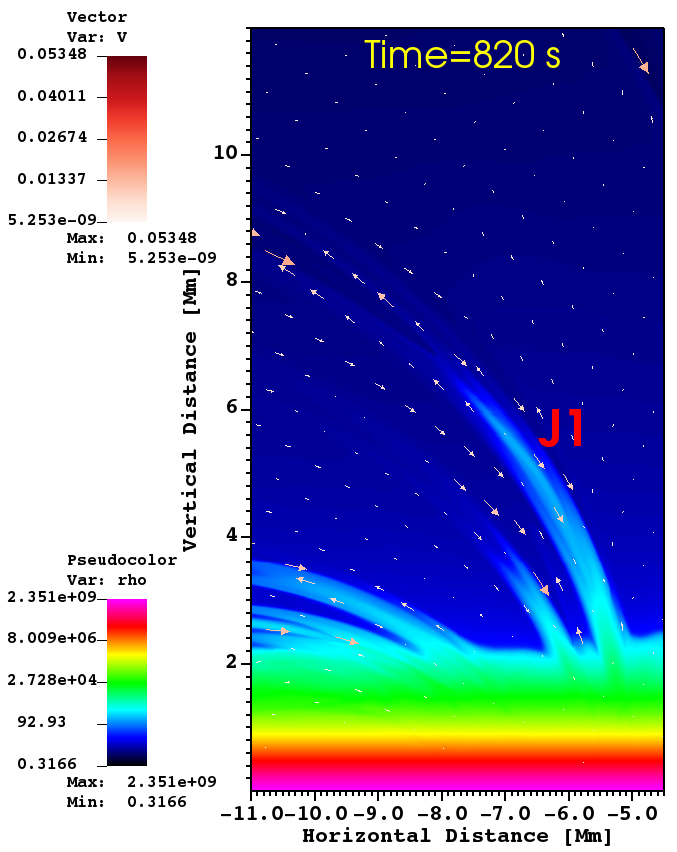}
}
\caption{The normalized density map of the evolution of cool 
jet J$_{1}$ as marked in Figure~\ref{fig}. This jet is triggered by the implementation of random vertical velocity, 
$V_{y}$, 
pulses. The pulses are applied in the chromosphere between 1 Mm  
and 1.5 Mm. The normalized velocity vectors are over-plotted on each density map between $t = 0$ s 
and 820 s. The time is in increasing order from top to bottom.}
\label{fig3} 
\end{figure}

 This jet evolves along the curved magnetic field lines 
(see 
Figures \ref{fig} and \ref{fig3}) and rises up to the height of 6.2 Mm above the TR. The plasma  (with different density streaks) flows along the curved field lines, forming the spine of the jet J$_{1}$  
(see 
Figure~\ref{fig3}). The multiple magnetoacoustic shocks interact with each other and with the null point region in the inner corona, which further creates  a complex velocity field locally in and around the evolved jets and plasma flows. Moreover, the jet's spine becomes {radially inhomogeneous}, with its core at higher density while the outer periphery is of a lesser density, as 
 one can see 
in Figure~\ref{fig}. 

\begin{figure}[H]   
\hspace{-2.0cm}
\mbox{\includegraphics[width=7.5cm,height=7cm]{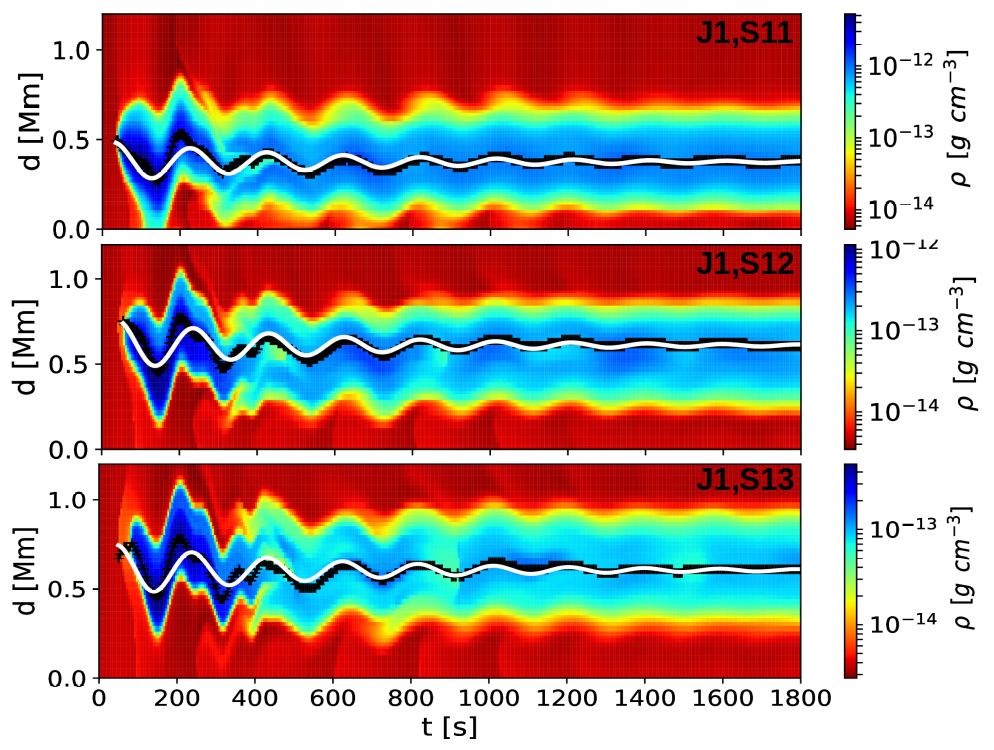}
\includegraphics[width=7.5cm,height=7cm]{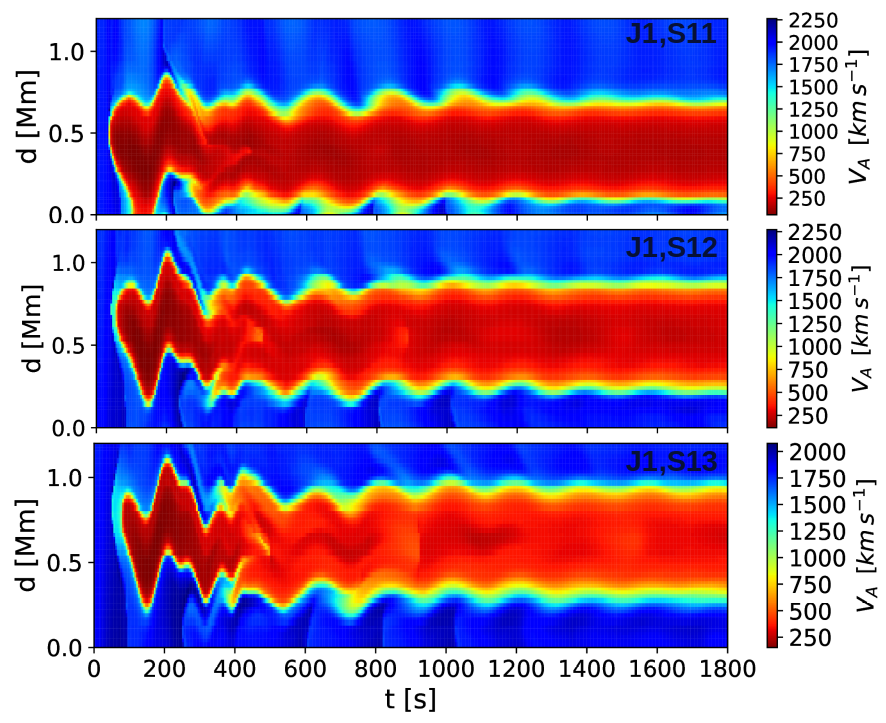}}
\caption{\textbf{Left:} 
the  
distance--time diagram of the density map for jet `J$_{1}$' corresponding to the slits S11, S12, and S13 
 from  
Figure~\ref{fig}. These density maps show the decaying transverse oscillations in jet 
J$_{1}$. 
The maps 
also display the cross-field structuring of density across the jet. \textbf{Right:} 
 the 
 distance--time diagram in Alfv\'en velocity space for jet `J$_{1}$' corresponding to the slits S11, S12, and S13 
 from  
Figure~\ref{fig}. These maps well 
 demonstrate the evolved cross-field structuring respectively in density and the Alfv\'en speed across the oscillating jet.}
\label{Fig:4}
\end{figure}

The evolved spicule-like cool jet J$_{1}$ is under the influence of the complex velocity fields locally created in the inner corona above the TR. This further triggers the transverse oscillations within 
the jet. 
This oscillation bends the axis of the jet as a whole, thus termed a transverse kink wave (or oscillation). To analyze the physical properties of the evolved transverse oscillations, we select slits S11, S12, S13 at jet J$_{1}$. 
Figure~\ref{Fig:4} shows the 
derived distance--time maps for the density (Figure~\ref{Fig:4}, left)  
and also in terms of Alfv\'en velocity space (Figure~\ref{Fig:4}, right) 
for J$_{1}$. 

Figure~\ref{Fig:4} shows the decaying transverse oscillations. The fitted decaying sinusoidal curve confirms the evolution of the decaying transverse
  waves in the first jet.  The fitting function is given as follows:
 \begin{equation}
     y(t) = b + A \exp({-\beta t}) \cos(\omega t + \phi)\,, 
\label{eq14}
 \end{equation}
where 
 $b$ denotes 
the background, 
 $A$  
represents the displacement amplitude, 
$\beta$ 
represents the damping coefficient in s$^{-1}$,  
$\omega$ 
 is the frequency of oscillation in (radian~$\cdot$~s$^{-1}$), 
and  
$\phi$ 
 denotes 
 the phase shift in radian.  The traverse oscillations can be manifested as kink oscillations, which are shown at three different 
heights along the jet  J$_{1}$, in 
 Figure~\ref{Fig:4}, left
top (for slit S11), middle (for slit S12), and bottom (for slit S13). 
The fitting parameters of these oscillations are given in Table \ref{Table:1}. The amplitudes of the oscillations lie in the range of  
114--155 km, while the damping time is of about 
550 s and the mean wave period is 
 about 
195 s. It should be noted that before 400 s, the transient oscillations are mixed with the natural transverse kink oscillations; however, 
 there is no  
need to fit the maps 
separately to extract the oscillations 
from the distance--time map in density. We comment on the complex nature of the oscillation in terms of the estimated velocity signals later. The highly dense J$_{1}$, which is triggered along the significantly curved magnetic field lines, supports the propagating transverse kink waves of 
period of about  
195 s at a speed of 
 about 
125 km s $^{-1}$. The propagation speed is derived by estimating the distance between slits S11 and S13, which is 
of approximately 
2 Mm, and the phase difference is 
 of approximately 
0.52 radian (see Table 1). The relation between the phase difference, period 195~s, and time delay is  $\Delta \phi= 360 ^{\circ}\times freq\times \Delta t$, where 
 $freq$ denotes 
the wave frequency as estimated from Table 1. 
 The phase difference of 0.52 radian gives almost 29.79$^{0}$, and the mean period 195 s gives the frequency 5.12 mHz.~Therefore, the time delay in the wave propagation 
between slits S11 and S13 is estimated approximately 16 s, and the phase speed in the high-density jet is calculated as about 125 km s$^{-1}$. The wavelength of the propagating kink wave is estimated to be as 
large as approximately  
 24,400 km. {Here, we compare the physical parameter (phase speed) of the wave derived from the analyzed data of full MHD simulations with the 
comparatively simpler analytical results.} By assuming a straight and slender {magnetized} flux tube, the kink 
speed, c$_{k}$,   
can also be expressed as follows 
\citep{1983SoPh...88..179E,2000SoPh..193..139R,2005LRSP....2....3N}:
   \begin{equation}
    c_{k} = \sqrt{\frac{\varrho_{\rm in}C_{A{\rm in}}^2+\varrho_{\rm ex}C_{A{\rm ex}}^2}{\varrho_{\rm in}+\varrho_{\rm ex}}},
 \end{equation}
  where $\varrho_{\rm in}$, $\varrho_{\rm ex}$, 
 $C_{A{\rm in}}$, $C_{A{\rm ex}}$ 
are, respectively, the internal and external densities and Alfv\'en speeds. In the core of J$_{1}$, the Alfv\'en speed is 
 around 
 125 km $^{-1}$, while, slightly outside towards jet's periphery, it is around 
 500 km $^{-1}$ 
(see 
  Figure~\ref{Fig:4}, right). 
 Similarly, the density contrast,  $\varrho_{\rm ex}/\varrho_{\rm in}$,  is almost 0.01 
 (see 
 Figure~\ref{Fig:4}, left).
~Therefore, 
the estimated kink speed, $c_{k}$, is found to be \mbox{about 
134~km s$^{-1}$}. The kink speed derived from the fitting parameters of the kink oscillations and time-lag analyses is consistent with the one derived by the MHD theory. 

\begin{figure}[H]   
\mbox{
\hspace{-2.0cm}
\includegraphics[width=8cm,height=4cm]{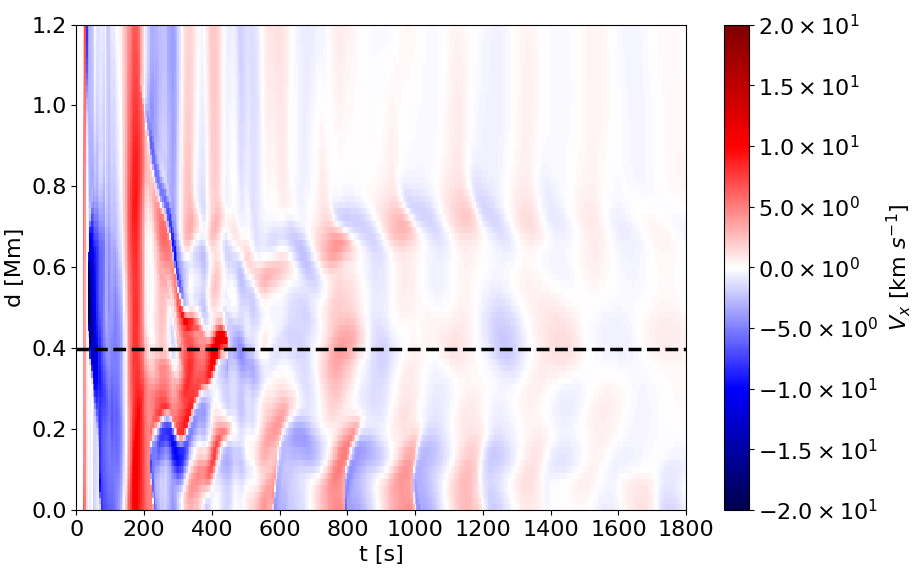}
\includegraphics[width=8cm,height=4cm]{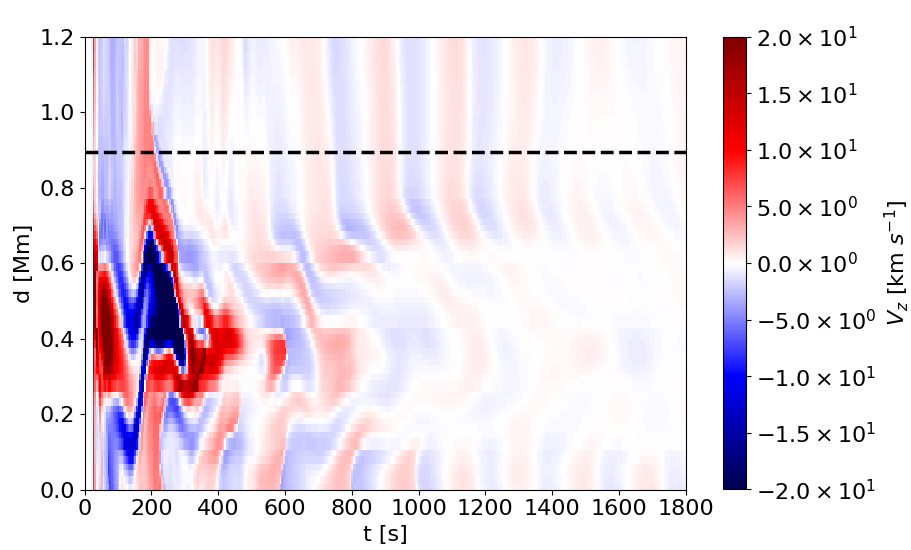}}
\includegraphics[width=8cm,height=4cm]{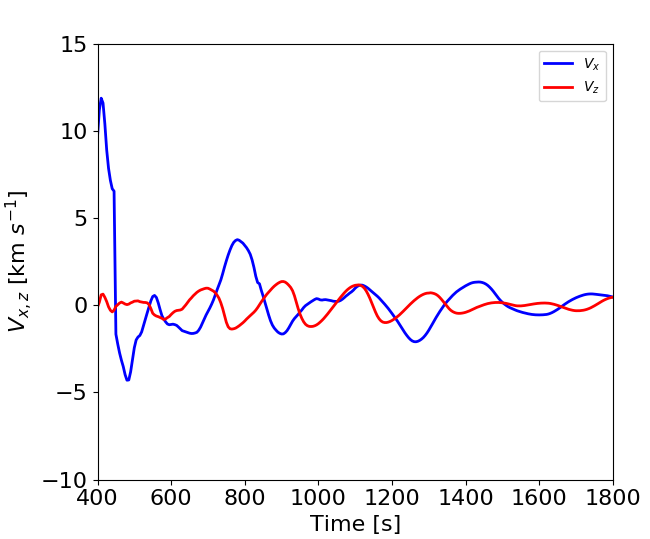}
\caption{The distance--time maps in 
$V_{x}$ (\textbf{top left}) 
and 
$V_{z}$ (\textbf{top right}) corresponding to slit S11 for the jet J$_{1}$ (see text for detail). 
 {\bf Bottom:} 
 a comparison of the V$_{x}$ signal derived from inside the jet
over the path, 
i.e., the dotted line 
 shown  
in the top 
 left panel, 
and the
$V_{z}$ signal estimated from the outer periphery of the jet in its ambient area over the path as shown 
by the dotted line 
in the top 
right panel.}
\label{Fig:5} 
\end{figure}

\begin{table}[H]
    \centering
    \caption{Fitting parameters of the transverse oscillations in jet J$_{1}$ 
(see Equation (\ref{eq14})). 
\label{Table:1}}

\setlength{\cellWidtha}{\textwidth/6-2\tabcolsep-0in}
\setlength{\cellWidthb}{\textwidth/6-2\tabcolsep-0in}
\setlength{\cellWidthc}{\textwidth/6-2\tabcolsep-0in}
\setlength{\cellWidthd}{\textwidth/6-2\tabcolsep-0in}
\setlength{\cellWidthe}{\textwidth/6-2\tabcolsep-0in}
\setlength{\cellWidthf}{\textwidth/6-2\tabcolsep-0in} 
\scalebox{1}[1]{\begin{tabularx}{\textwidth}{>{\centering\arraybackslash}m{\cellWidtha}>{\centering\arraybackslash}m{\cellWidthb}>{\centering\arraybackslash}m{\cellWidthc}>{\centering\arraybackslash}m{\cellWidthd}>{\centering\arraybackslash}m{\cellWidthe}>{\centering\arraybackslash}m{\cellWidthf}}
\toprule
         \multirow{2}{*}{\textbf{Jet} }& \multirow{2}{*}{\textbf{Slit}}& 
{\boldmath$A$}
& 
{\boldmath$\beta$} 
& 
\boldmath{$\omega$} 
&
\boldmath{$\phi$} 
\\
         
         & & \textbf{(Mm)}  & \textbf{(s}\boldmath{$^{-1}$}\textbf{)} &\textbf{ (rad s}\boldmath{$^{-1}$}\textbf{)} & \textbf{(rad)}  \\
         
         \midrule
         
         \multirow{6}{*}{J$_{1}$} & S11 & 0.1143 $\pm$ 
0.0067 & 0.0018 $\pm$ 
0.0001 & 0.0318 $\pm$ 
0.0001 & $-$0.95857 
$\pm$ 
0.0582 \\
                  \cmidrule{2-6}
         & S12 & 0.1557 $\pm$ 
0.0099 & 0.0018 $\pm$ 
0.0001 & 0.0323 $\pm$ 
0.0001 & $-$1.3568 $\pm$ 
0.0596  \\
                  \cmidrule{2-6}
         & S13 & 0.1545 $\pm$ 
0.0117 & 0.0018 $\pm$ 
0.0002 & 0.0325 $\pm$ 
0.0002 & $-$1.4839 $\pm$ 
0.0729 \\
         
         \bottomrule
    \end{tabularx}}
\end{table}
Figure~\ref{Fig:5}, top left,  
displays the distance--time map for the 
$V_{x}$-component  
of the velocity along slit S11, while the {top-right} panel shows the distance--time map for the 
$V_{z}$-component of the velocity. 
 One can see
that before 400s,
strong Alfv\'enic motions are set into the jet, where both the $x$- 
and 
$z$-components 
of the velocities are present, indicating the evolution of this mixed mode, where the axial ($x$-component 
representing kink oscillations in Cartesian coordinate system) and $z$-component (representing Alfv\'en motions in 
2.5D 
regime) of velocity evolve together. Between 400 s 
and 1200 s, when these transverse oscillations decay, 
as well 
seen in the distance--time maps of density (Figure~\ref{Fig:4}, left), 
in the 
 $V_{x}$ 
distance--time map, the periodic velocity 
($V_{x}$) 
oscillations {dampen} in magnitude within the jet. 
 One can alos see 
that in the same time domain, such velocities are weakly present in the ambient plasma outside the jet. Co-temporally, in the 
 $V_{z}$ 
distance--time map, the periodic velocity 
 ($V_{z}$)  
oscillations are enhanced in the ambient plasma outside and at the boundary layers of the jet, while it is weakly present within the jet. After 1200 s, 
there is almost no transverse oscillation that remains within the jet.~{This scenario is 
 visible  
in 
Figure~\ref{Fig:5}, bottom. 
 This physical scenario 
is compared with the 
 $V_{x}$  
 signal derived from inside the jet over the path, 
i.e., the dotted line, 
as shown in the Figure~\ref{Fig:5}, 
 top left,  
and the 
 $V_{z}$ signal estimated from the outer 
periphery of the jet from its ambient area over the path, as displayed by the dotted line shown in 
Figure~\ref{Fig:5}, top right. 
It is 
 understood 
that the 
 $V_{x}$ 
velocity amplitude oscillations dampen between 400 s  
and 1200 s, while the 
 $V_{z}$  
velocity amplitude oscillations amplify during the same time period. Beyond this time, there is no such correlation that exists, 
suggesting that the resonant absorption is stopped due to the continuously changing physical conditions of the 
dynamic jet \citep{2011SSRv..158..289G,2018A&A...619A.173S}.}

\begin{figure}[H]
\begin{adjustwidth}{-\extralength}{0cm}
\centering 
\mbox{\includegraphics[width=5cm,height=5cm]{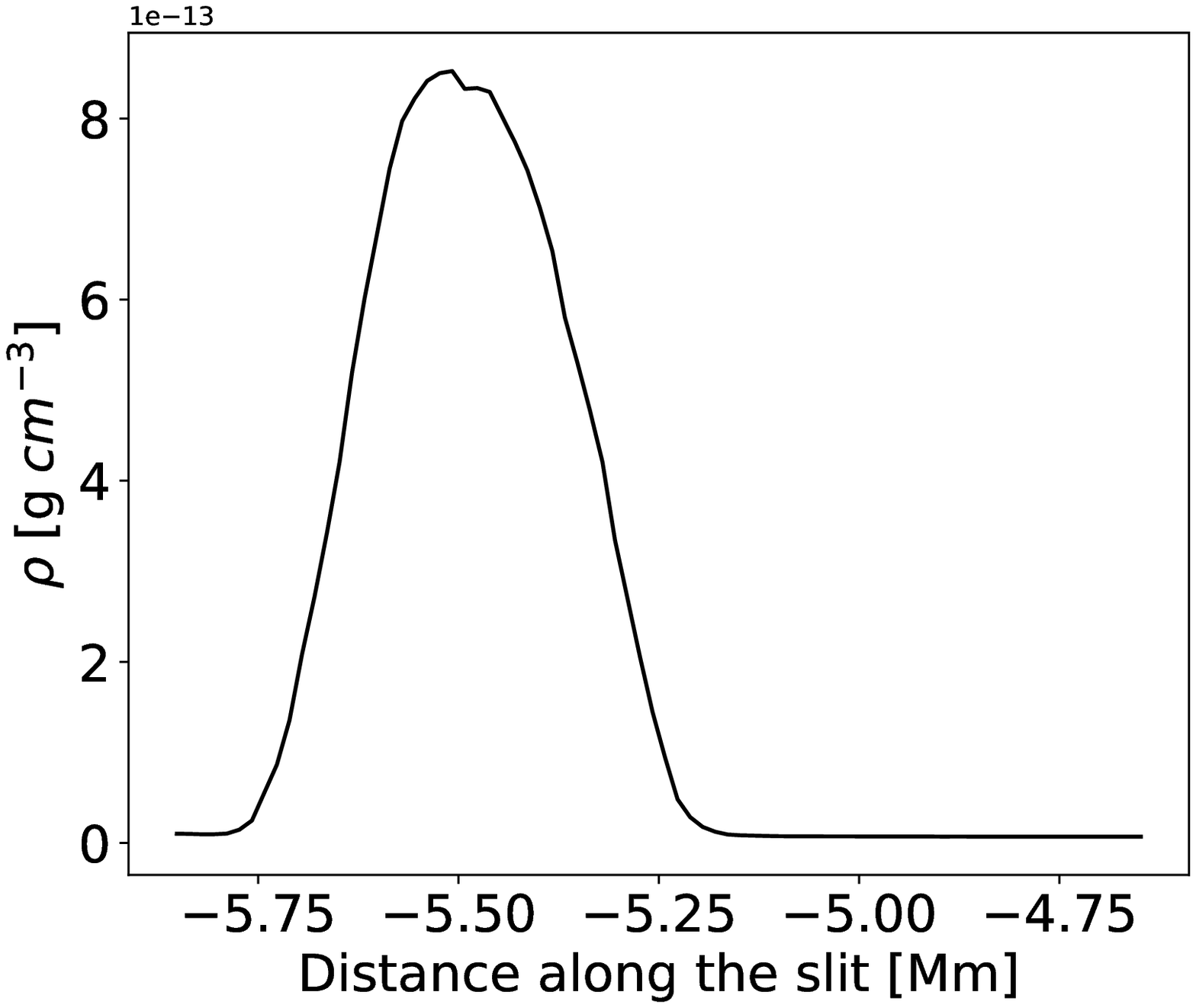}
\includegraphics[width=5cm,height=5cm]{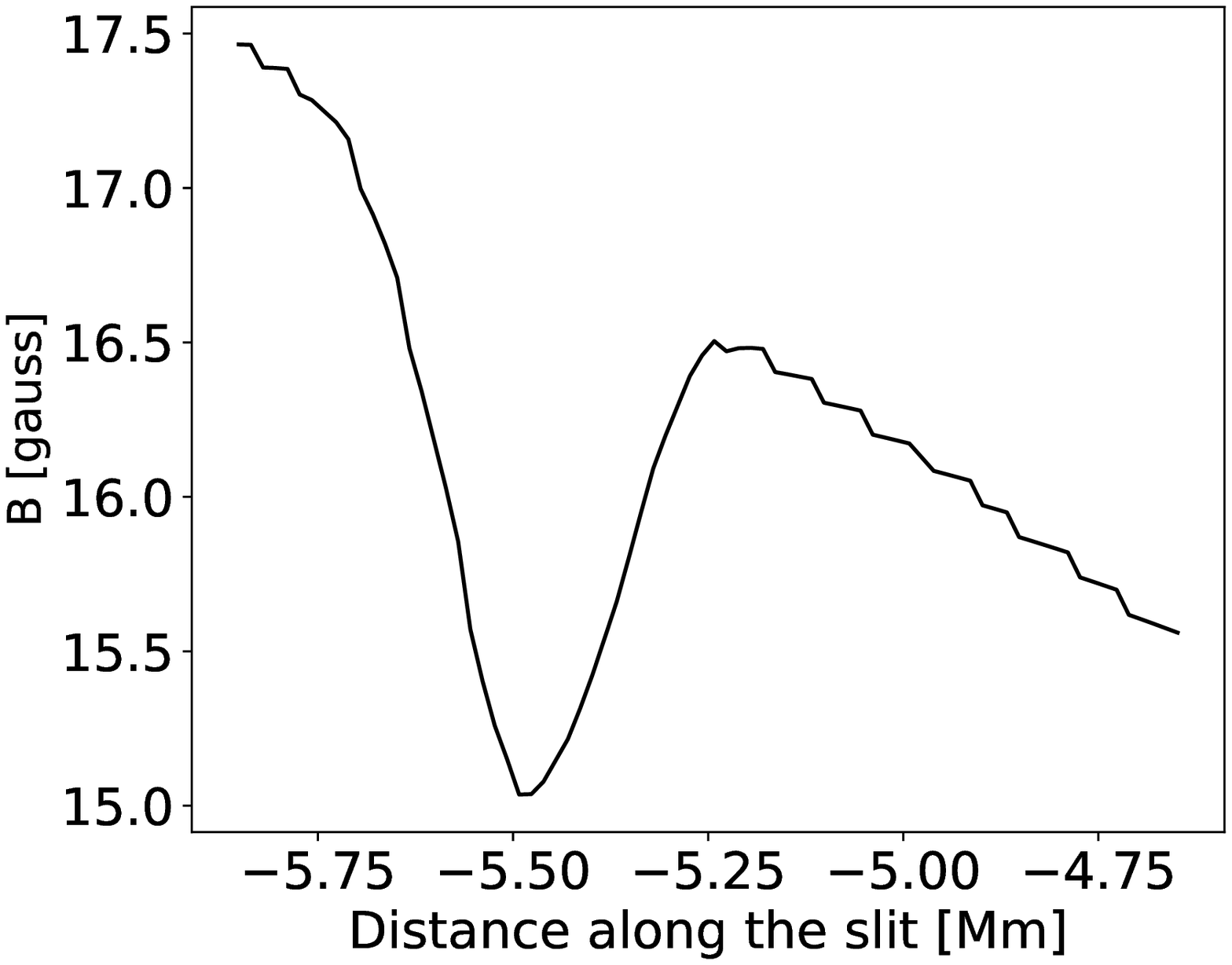}
\includegraphics[width=5cm,height=5cm]{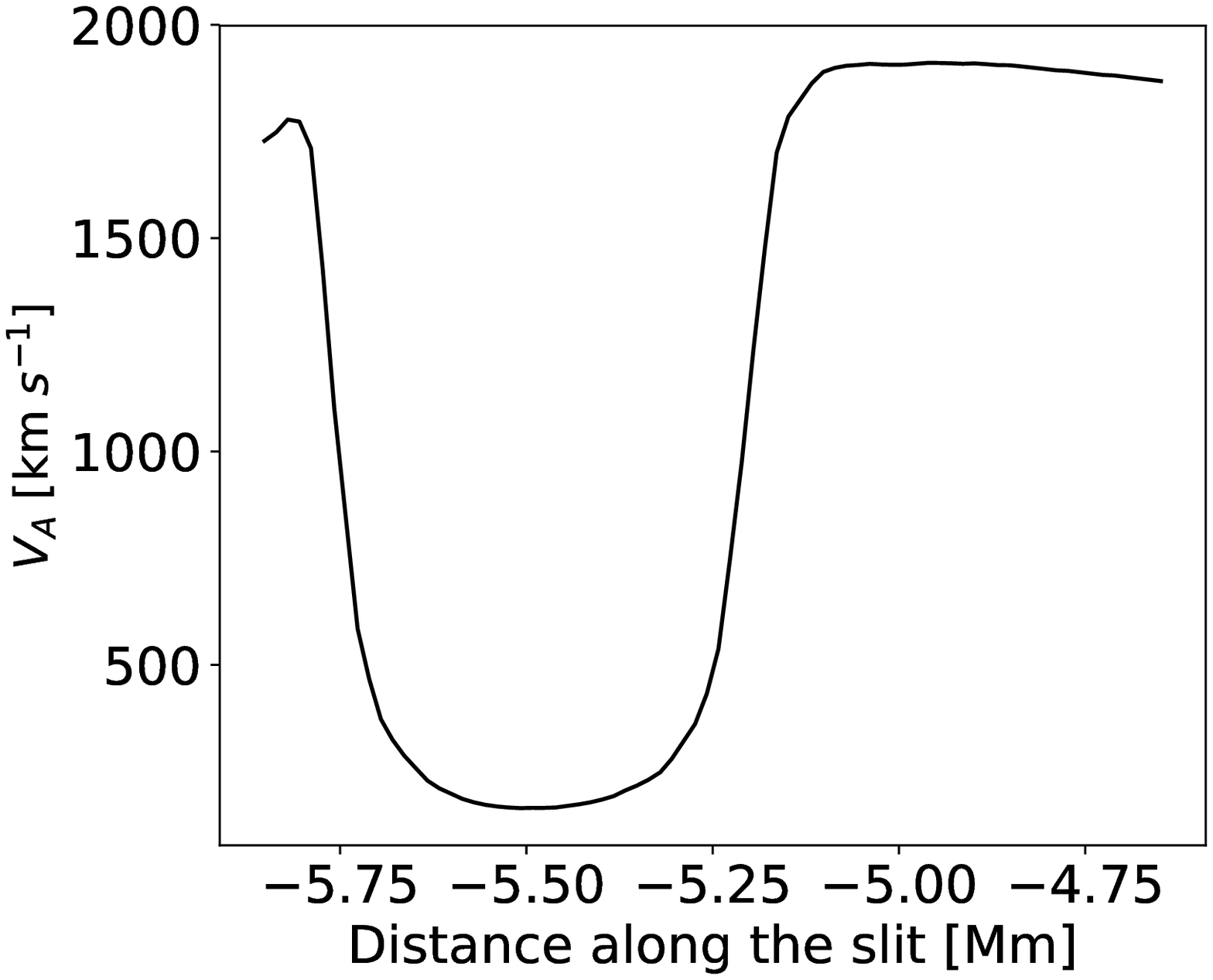}}
\end{adjustwidth}
\caption{
The density ({\bf left}),   
magnetic field ({\bf middle}), and 
 Alfv\'en speed ({\bf right}) versus 
distance along slit S11 for Jet J$_{1}$ at 
$t=1600$ s. 
See text for details. 
}
\label{fig6}
\end{figure}

Figure~\ref{fig6} , 
represents the density 
(Figure~\ref{fig6}, left),  
 the magnetic field  
(Figure~\ref{fig6}, middle), and the 
Alfv\'en speed 
(Figure~\ref{fig6}, right) 
 versus  
length along slit S11 at $t=1600$ s 
for the J$_{1}$ jet. This shows that when 
 one moves   
outward across the jet, the density decreases, while the magnetic field and Alfv\'en speed 
increase. We show that the cross-field structuring of the density and characteristic Alfv\'en speed across the jet spine is created when the jet is evolving and oscillatory motions appear there.  We conjecture that the cross-field structuring of the density and characteristic Alfv\'en speed across the jet J$_{1}$, as well as the curvature of the magnetic field, causes the onset of the leakage/conversion of wave energy outward to dissipate these kink waves (in case of J$_{1}$). The kink oscillation decays smoothly in J$_{1}$ at each height (S11--S13). There is no intrinsic decay mechanism incorporated in this ideal MHD simulation {to deal with the damping of the wave by cooling agents or collisional dissipative properties (e.g., viscosity, etc.).}
As stated in the above paragraph, the variations in the 
$V_{x}$ and 
 $V_{z}$  
provide the true scenario of wave {damping}. The resonant absorption sets in due to the cross-field structuring of the density and Alfv\'en speed across the jet. It causes the {damping} of kink oscillations (associated with 
 the 
 $V_{x}$-component  
of velocity), and, simultaneously, at the boundary layer of the jet and in the ambient plasma, the Alfv\'enic motions (associated with the 
 $V_{z}$-component  
of the velocity) are enhanced. 


The evolved kink wave has a phase speed of 125 km s$^{-1}$ and the maximum velocity amplitude of about 
$\pm\,$5 km s$^{-1}$. 
The average energy flux of the kink wave is 
$F=0.5f(\varrho_{\rm in}+\varrho_{\rm ex}){V}^{2}V_{p}$~\citep{2014ApJ...795...18V}. 
 {Here,}
$V$ 
 and 
$V_{p}$ 
are 
the  
velocity amplitude and 
phase velocity, respectively, and $f$ is the filling factor. 
~By~considering the  
$\varrho_{\rm in}+\varrho_{\rm ex}$ 
$\approx$ 10$^{-12}$ g cm$^{-3}$, $V$ $\approx$ \mbox{$\pm$4 km s$^{-1}$}, $V_{p}$ = \mbox{125 km s$^{-1}$,} the 
energy flux of the kink wave is obtained 
 to be about 
 1.0 $\times$ 10$^{6}$ \mbox{ergs cm$^{-2}$ s$^{-1}$} for the filling factor 1.
~This energy, if dissipated through the resonant absorption 
into the coronal ambient area where the jet is propelled, is sufficient for the coronal heating.~{The radius or half-width, 
$R$, 
of the jet is about} 
200 km.~In Figure~\ref{Fig:5},  
left, 
the 
$V_{x}$ 
velocity amplitude oscillation has a discontinuity (drop in a 
 considerably  
thin localized region almost to zero), as 
 well visible  
in the line around $d$ = 0.6 Mm and $t$ = 600--1200 s duration, 
 as soon as one moves  
out across the jet. In a similar manner, there is a drop in the 
 $V_{z}$  
velocity amplitude oscillation in the same region. This behavior is consistent with previous analytical models (e.g., \cite{2016ApJ...831...30Y}). The width, 
$l$, 
 of this transition region is 
  considerably 
thin, i.e., approximately 
30 km. Therefore, the $l$/$R$-ratio, in the present case, is approximately 
 0.15, which is consistent with 
various previous analytical estimations demonstrating strong resonant absorption \citep{2012A&A...546A..82S,2011SSRv..158..289G,2018A&A...619A.173S}.
       
\section{Discussion and Conclusions}\label{sec:4}

The transverse oscillations in spicule or spicule-like structures are detected in an abundant measure in the observational baseline \citep{article,2007A&A...474..627Z,2007Sci...318.1574D,Jess_2011,2018ApJ...853...61S,2021ApJ...921...30S,2022ApJ...930..129B}. \citet{Jess_2011} observed the transverse oscillations in various spicules, which were decaying kink waves. These observations are 
 quite 
similar to the modeled oscillations in the present paper. The jet J$_{1}$ is propelled up into the inner corona, carrying the cool and denser plasma. It is also subjected to kink oscillations. Over a longer time period, this jet is also subjected to the plasma flows along the spine. This creates variations in the density, magnetic field, and Alfv\'en speed across this jet. Moreover, they are formed on the spines of  curved magnetic field lines.

Due to this inhomogeneity and the presence of an inhomogeneous transitional 
layer in transverse direction, the kink waves are damped spatially by the resonant absorption 
(e.g., \cite{1978ApJ...226..650I,1988JGR....93.5423H,1991SoPh..133..227S,1992SoPh..138..233G,2002A&A...394L..39G,2006RSPTA.364..433G,2009A&A...503..213G}
and  references 
 therein).
In the location within the inhomogeneous transitional layer at which the wave frequency matches the local Alfv\'en frequency, i.e., the resonance position, the kink wave's properties change and it becomes more Alfv\'enic as the wave propagates (see Figure~\ref{Fig:5}). Due to this physical process, the transverse kink motions of the flowing flux tube (for example, the jet here) are damped, while the Alfv\'enic motions within this transitional layer (here, the jet's boundary and the ambient plasma outside) are amplified.

In the jet (J$_{1}$) that is triggered along significantly curved magnetic field lines, the propagating transverse kink wave of period 195 s at a speed of 125 km s$^{-1}$ is generated. 
We conclude that the cross-field structuring of the density and characteristic Alfvén speed (Figures~\ref{Fig:4} and \ref{fig6}) within J$_{1}$ most likely cause the onset of resonant absorption to dissipate the kink waves. 
This also implies the resonant energy transfer of transverse waves outward. A fraction of the {damping} of kink waves can also occur due to the wave leakage by the curved magnetic fields of the jets 
\cite{2004psci.book.....A}.  
 However, 
 the contribution of wave leakage
may be considered lower than the resonant absorption. 

The transverse oscillations and Alfv\'en waves are modeled in the lower solar atmosphere, producing solar spicules or spicule-like jets. Spicules may be generated due to the amplification of the magnetic tension transported upward through interactions between ions and neutrals or ambipolar diffusion in the solar chromosphere, which drives flows, heat, and Alfv\'enic waves \cite{2017Sci...356.1269M}. The spicular-type structures are associated with torsional Alfv\'en waves \cite{2017NatSR...743147S}. Random motions in the solar photosphere and chromosphere can generate Alfv\'en waves, which non-linearly couple with the longitudinal motions and generate the spicule-like plasma ejecta \cite{1982SoPh...75...35H,1999ApJ...514..493K,10.1093/mnras/stac252}. \citet{2016ApJ...829...80B} reported that the ponderomotive coupling from Alfv\'en and kink waves into slow modes generates shocks, which can have a capacity to heat the upper chromosphere, and may drive solar chromospheric spicules. Most of these previous numerical models advocate for either Alfv\'en (and/or kink) waves as a driver of solar spicules or the evolution of these transverse waves in the static spicular flux tubes. In the present model, we firstly study a physical scenario in which impulsive multiple velocity perturbations in the localized chromosphere may launch both the spicule-like cool jets and the transverse oscillations in some of them. This physical scenario is 
close enough 
to the complex plasma motions seen originating from the solar chromosphere (i.e., multiple cool jets, their quasi-periodic rise and fall, oscillations, flows in closed field lines, etc.). Moreover, we demonstrate a compelling signature of the onset of resonant absorption as a primary {damping} mechanism of the transverse oscillations excited in a model jet. These kink waves have sufficient energy flux, which, when dissipated through the resonant absorption, is sufficient for the heating of the ambient coronal plasma.
A more detailed parametric study of these oscillating jets,  the physical nature of their {damping} mechanisms, and seismological aspects will be presented in a forthcoming 
 study. 

\authorcontributions{
Conceptualization, A.K.S.; methodology, A.K.S. and B.S.; software-run, B.S.; validation, A.K.S. and B.S.; formal analysis, B.S.; investigation, A.K.S. and B.S.; resources-utilized, A.K.S. and B.S.; data curation, B.S.; writing---original draft preparation, A.K.S.; writing---review and editing, A.K.S. and B.S.; visualization, B.S.; supervision, A.K.S.; project administration, A.K.S.; funding acquisition, A.K.S. All authors have read and agreed to the published version of the manuscript.

\funding{A.K.S. 
acknowledges the ISRO Project Grant number DS\_2B-13-12(2)/26/2022-Scc2 
 for the support of his research. 
B.S. gratefully acknowledges the Human Resource Development Group (HRDG), Council of Scientific $\&$ Industrial Research (CSIR), India, for providing him with a senior research scholar grant. 

}

This research received no external funding.}

\dataavailability{The numerical simulation data are available and can be provided by the corresponding author on reasonable request.}

\acknowledgments{{The authors gratefully acknowledge all four reviewers for their very detailed and valuable remarks that improved our manuscript considerably.} 
B.S. acknowledges the National Supercomputing Mission (NSM) for providing computing resources for ‘PARAM Shivay’ at the Indian Institute of Technology 
 BHU), 
Varanasi, which is implemented by C-DAC 
and supported by the Ministry of Electronics and Information Technology (MeitY) and Department of Science and Technology (DST), Government of India. 
B.S. also acknowledges the use of the PLUTO code in the present work and the use of the PYTHON libraries for numerical data analysis. AKS and BS acknowledge the initial discussions from K. Murawski, J.J. Gonz{\'a}lez-Avil{\'e}s, and Yuan Ding. 
The authors thank the Guest Editors for their invitation to submit an article to the Special Issue in the Honor of Professor Marcel Goossens on the occasion of his 75th birthday. 
We gratefully acknowledge the pioneering contributions of Prof. M. Goossens in magnetohydrodynamics (MHD) and in particular his paramount contributions on the MHD waves and their  dissipation due to resonant absorption, which all led a significant development in the existing knowledge of the current generation working in the area of solar and astrophysical plasma as well as wave theory.}

\conflictsofinterest{The authors declare no conflict of interest.}

\begin{adjustwidth}{-\extralength}{0cm}

\reftitle{References}
\PublishersNote{}
\end{adjustwidth}
\end{document}